\documentclass[10pt]{article}

\usepackage{authblk}
\usepackage{amsmath,bm}
\usepackage{graphicx}
\usepackage{subfigure}

\title{Volumetric Lattice Boltzmann Models in General Curvilinear Coordinates: Theoretical Formulation}

\author{Hudong Chen}
\affil{Dassault Systemes, 175 Wyman Street, Waltham, MA 02451}

\begin{document}

\maketitle


\begin{abstract}
A theoretical formulation of lattice Boltzmann models on a general curvilinear coordinate system is presented.  
It is based on a volumetric representation so that mass and momentum are exactly conserved as in the conventional lattice  
Boltzmann on a Cartesian lattice.  In contrast to some previously existing approaches for arbitrary meshes involving interpolation approximations among multiple neighboring cells, the current formulation preserves the fundamental one-to-one advection feature of a standard lattice Boltzmann method on a uniform Cartesian lattice.  
The new approach is built on the concept that a particle is moving along a curved path. A discrete space-time inertial force is derived so that the momentum conservation is exactly ensured for the underlying Euclidean space. We theoretically show that the new scheme recovers the Navier-Stokes equation in general curvilinear coordinates in the hydrodynamic limit, along with the correct mass continuity equation.
   
\end{abstract}



\section{Introduction}

Lattice Boltzmann Methods (LBM) have been developed as an advantageous method for computational fluid dynamics during past few decades.\cite{Benzi,ChenDoolen} The underlying dynamics of a lattice Boltzmann system resides on the kinetic theory, in that it involves motion of many particles according to the Boltzmann equation.\cite{Huang,Cerci}  
There are two fundamental dynamical processes in a basic Boltzmann kinetic system, namely advection and collision. 
In the advection process, a particle moves from one location to another according to its microscopic velocity, while in the collision process particles interact with each other obeying conservation laws and relax to equilibrium. 
In a standard LBM model, particle velocity takes on a discrete set of constant values, and the latter form exact links from one lattice site to its neighboring sites on a simple Bravais lattice corresponding to a three dimensional (3D) uniform cubical Cartesian mesh.\cite{Qian,CGO,Shan} Therefore, having these discrete velocity values, a particle moves from one lattice site to a unique neighboring site per constant time interval (say $\Delta t = 1$). 
This exact one-to-one advection of a particle on the lattice effectively realizes a Lagrangian advection characteristics. That is, a particle location during the advection process is exactly predicted at any later time from its location at a given time. 
Such a Lagrangian advection characteristics is the reason for one of the key advantages of LBM, besides its simplicity, the advection process produces an extremely low numerical dissipation 
(due to no smearing of locations of a single particle). 
Indeed, if such a one-to-one advection is relaxed, numerical dissipation has been seen to significantly increase even with the usage of much higher order interpolation schemes. In addition, the Lagrangian one-to-one advection may also have non-trivial implications in simulations of flows at finite Knudsen numbers.\cite{Shan,jhon,zhang}
In contrast, unlike particle velocity, a fluid velocity varies in both time and space, hence an one-to-one advection process is impossible for a Navier-Stokes equation based numerical scheme on a spatially fixed mesh.
  
However, a uniform and cubical Cartesian mesh poses fundamental limitations. 
First of all, often in realistic fluid study one deals with a solid geometry with curved surfaces, and obviously a Cartesian mesh does not smoothly conform with such a geometric shape. 
Secondly, a flow usually has small scale structures concentrated in certain spatial locations and directions. 
For instance, in the turbulent boundary layer, the flow scale in the direction normal to the wall is much smaller than in the tangential direction or in the fluid region outside of the boundary layer. 
Consequently, the requirement on spatial resolution is significantly higher in the normal direction inside a boundary layer. A cubic Cartesian mesh does not provide the flexibility with different spatial resolutions in different directions. Therefore, it is of fundamental importance to extend the standard LBM for a general non-cubic Cartesian mesh.

There have been various attempts in the past to extend LBM on arbitrary meshes, including curvilinear mesh (cf. \cite{He,volumetric,Barraza}).  
The essential idea is to relax the aforementioned exact one-to-one advection between a pair of lattice sites while keeping the advection of particles according to the original constant discrete velocity values. As a consequence, on an arbitrary mesh, a particle after advecting from its original mesh site does not in general land on a single neighboring mesh site. Thus its location needs to be distributed onto a set of neighboring mesh sites via interpolation. As mentioned above, such an effective ``one-to-many'' advection process destroys the preciseness of a Lagrangian advection characteristics thus resulting in a significantly increased numerical dissipation.

It is of fundamental interest to formulate an LBM on a general mesh that preserves the exact one-to-one feature of the particle advection process. A possible way to accomplish this is to construct the micro-dynamic process on a non-Euclidean space represented by a general curvilinear coordinate system based on Riemann geometry.\cite{book} In such a non-Euclidean space, a constant particle velocity corresponds to a curved and spatially varying path in the underlying Euclidean space. Indeed, the continuum Boltzmann kinetic theory on a Riemannian manifold can be theoretically described.\cite{veden,love} The key difference between a constant particle velocity on a Euclidean and a non-Euclidean space is that the latter is accompanied with an inertial force, due to the 1st law of Newton in Euclidean space.  
According to this concept, it is entirely conceivable to formulate LBM models on a cubic Cartesian lattice in non-Euclidean space that corresponds to a general curvilinear mesh in the Euclidean space.
The first such attempts were made by Mendosa et al. in a series of papers (see, \cite{Mendosa1,Mendosa2,Mendosa3}).  
The significance of their work is that it established this fundamental approach to 
LBM on curvilinear meshes via a Riemann geometric framework.
  
As Mendosa et al., we formulate LBM on a general curvilinear coordinate system via the concept of
Riemannian geometry. There are several critical differences between our approach presented here and that of Mendosa et al.  
First of all, we adopt a volumetric formulation so that the mass and momentum conservations are exactly ensured. The resulting continuity equation of mass is automatically shown to have the correct form in curvilinear coordinates, and there is no need to introduce any artificial mass source term to correct for any artifacts in the resulting hydrodynamics.  Secondly, as in the continuum kinetic theory on a manifold, the only external source term in the extended LBM is associated with an effective inertial force due to curvilinearity. It contributes no mass source.
Moreover, such a force term is constructed fully self-consistently within the discrete lattice formulation. 
This force form in discrete space and time recovers asymptotically the one in the continuum kinetic theory in the hydrodynamic limit. 
It does not rely on any outside analytical forms borrowed from the continuum kinetic theory.\cite{veden,love} 
Thirdly, the inertial force enforces the exact momentum conservation for the underlying Euclidean space in the discrete space-time LBM model.
Lastly, we demonstrate that the force term must be constructed properly so that it adds momentum in the system at a proper discrete time moment in order to produce the correct resulting Navier-Stokes hydrodynamics to the viscous order. Indeed, the correct Navier-Stokes equation is fully recovered in the hydrodynamic limit with the new formulation without introducing any extra correction terms.

To summarize, our goal is to theoretically demonstrate how discrete kinetic theories can be formulated in curvilinear coordinates in a way that mass and momentum are exactly conserved, and also the correct hydrodynamics is recovered in the microscopic limit. We hope the reader may find such discrete non-equilibrium statistical systems interesting, similar to discrete systems found in other branches of physics, e.g. spin models on a lattice in the equilibrium statistical theory, lattice field theories, etc. In addition to fundamental interest, non-equilibrium discrete systems like LBM may be useful for applications in numerical analysis and for practical use. Computational aspects such as scheme stability, numerical dissipation, performance, and boundary conditions treatment, as well as numerical verification of theoretical results, are outside of the scope of this study. We also restrict ourselves to one of the simplest situations, namely the low speed isothermal flow, which admits a closed self-contained analytical treatment which explicates the main features of volumetric formulation in curvilinear coordinates. Possible extensions and applications of this work are briefly discussed in Section 4 below. 

The description of non-equilibrium dynamics and transport at the spatio-temporal scales not treatable by hydrodynamics can often benefit from using kinetic theory methods. Here we show how to construct lattice kinetic theories in a general coordinate system that possess exact conservation laws at both kinetic and continuum scales, as well as correct macroscopic limit behavior. In addition to the fundamental interest, this can be useful for describing interactions at interfaces between different fluids or materials with complex geometry.

The subsequent sections are organized as follows. We describe in Section 2 the formulation of the new LBM on a general curvilinear mesh. In Section 3, we provide a detailed theoretical derivation and show that the new LBM indeed produces the correct Navier-Stokes equation in general curvilinear coordinates. In order for the paper to be more self-contained, we provide, in Appendices A, B and C, a basic theoretical description of the continuum Boltzmann kinetic theory in curvilinear coordinates as in the literature,\cite{love} some fundamental properties,\cite{book} as well as a derivation of the Navier-Stokes hydrodynamic equation on a general curvilinear coordinate system.

\section{Formulation of Lattice Boltzmann in Curvilinear Coordinates}        

This section is divided into two subsections.  First, we construct the geometric quantities that are necessary for defining a general curvilinear coordinate system.  Then we present the volumetric lattice Boltzmann formulation on a general curvilinear mesh.

\subsection{Construction of geometric quantities on a curvilinear mesh}

We describe how to construct a coordinate system for formulating kinetic theory on a curvilinear mesh. 
Let ${\bf x}$ be any spatial point in 3D Euclidean space.  
We choose a curvilinear coordinate system, so that ${\bf x} = {\bf x}({q})$ is uniquely defined by ${q}$. 
Here ${q} \equiv (q^1, q^2, q^3)$ are the coordinate values along three non-collinear congruences of parameterized basis curves.  
The above forms the basic starting point for a general curvilinear mesh. 
For when a 3D curvilinear mesh with its entire spatial layout is given, then all its vertex locations in space are known and specified. Let ${\bf x}$ now be only defined on sites (i.e., vertices) of the curvilinear mesh.
We choose the coordinate system $\{ {q} \}$ that is a one-to-one mapping between ${\bf x}$ and ${q}$. In other words, for any site ${\bf x}$ on the curvilinear mesh, there is a unique value ${q}$ associated with it, that is
${\bf x} = {\bf x}({q})$ is uniquely defined. In addition, we construct the coordinate values of ${q}$ on the mesh  as follows: For any site ${\bf x}({q})$, its nearest neighbor site along the $i$th ($i = 1, 2, 3$) coordinate curve in the positive or negative direction is a spatial point ${\bf x}_{\pm i}$.  
Due to the unique mapping, we have ${\bf x}_{\pm i} = {\bf x}({q}_{\pm i})$, where ${q}_{\pm i}$ is a unique coordinate value for the neighboring site.
It is entirely possible to choose the spatial variation of ${q}$ in such a way that ${q}_{\pm i}$ and ${q}$ only differ in their $i$th coordinate component values by a constant $d_0$.
That is, ${q}_{\pm i} = (q^1_{\pm i}, q^2_{\pm i}, q^3_{\pm i})$, 
and $q^j_{\pm i} - q^j = \pm d_0 \delta^j_i$ ($i,j = 1, 2, 3$). 
Here we choose the constant $d_0$ to be unity without loss of generality.

It is easy to see that, under such a construction, the coordinate values $\{ {q} \}$ actually form a simple uniform 3D cubic Cartesian lattice with the lattice spacing unity. 
One may interpret this ``Cartesian'' lattice $\{ {q} \}$ as a result of deformation (bending, twisting 
and stretching/compressing) of the original curvilinear mesh $\{ {\bf x} \}$ in the Euclidean space.  
Thus, its topological structure is the same as the original curvilinear mesh, but
the Cartesian lattice is on a non-Euclidean space as a result of distortion of the original Euclidean space.  
 
When a curvilinear mesh is provided, spatial locations of all the vertices $\{ {\bf x} \}$ on the
mesh are specified, and hence the distance from any one vertex to another on a such mesh is also fully determined. 
Let us define the distance vector from ${\bf x}({q})$ to one of its neighbors ${\bf x}({q}_{\pm i})$ ($i = 1, 2, 3$) as 
\begin{equation} 
{\bf D}_{\pm i} ({q}) \equiv {\bf x}({q}_{\pm i}) - {\bf x}({q}); \;\;\; i = 1, 2, 3
\label{distance}
\end{equation} 
For instance, ${\bf D}_{\pm 1} ({q}) \equiv {\bf x}(q^1 \pm 1, q^2, q^3) - {\bf x}(q^1, q^2, q^3)$.
Notice that, due to spatial non-uniformity of a general curvilinear mesh, the spatial distance from one mesh
site to its nearest neighbor site is in general  changing from location to location.
In other words, ${\bf D}_{\pm i} ({q})$ is a function of ${q}$.  
Furthermore, one can realize that the distance value in the positive direction along the $i$th coordinate curve
is in general not equal to that in the negative direction.  Explicitly, in terms of the distance vectors,
${\bf D}_i ({q}) \neq - {\bf D}_{-i} ({q})$.  For example, according to the definition
of (\ref{distance}),
\begin{eqnarray}
{\bf D}_{-1}({q}) &=& {\bf x} (q^1 - 1, q^2, q^3) - {\bf x} (q^1, q^2, q^3) 
= - {\bf D}_1 (q^1 - 1, q^2, q^3) \nonumber \\
&\neq& - {\bf D}_1 ({q}) = - ({\bf x} (q^1 + 1, q^2, q^3) - {\bf x} (q^1, q^2, q^3)) 
\end{eqnarray}
and the inequality only becomes an equality everywhere if the curvilinear mesh is a uniform lattice 
(so that $|{\bf D}_i| = const$, independent of spatial coordinate value ${q}$). 

Following the basic differential geometry concept, 
we now construct the basis tangent vectors at ${\bf x}({q})$ along each of the three coordinate directions, 
\begin{equation} 
{\bf g}_i ({q}) \equiv \beta_i ({q}) [{\bf D}_i ({q}) - {\bf D}_{-i}({q})]/2 \Delta x; \;\;\; i = 1, 2, 3 
\label{tangent}
\end{equation}
where the scalar factor $\beta_i ({q})$ is 
\[ \beta_i ({q}) \equiv [|{\bf D}_i ({q})| + |{\bf D}_{-i}({q})|] /|{\bf D}_i ({q}) - {\bf D}_{-i}({q})|\]
and $\beta_i ({q}) = 1$ when the two distance vectors ${\bf D}_i ({q})$ and $- {\bf D}_{-i}({q})$ are parallel. 
With such a construction, the parity symmetry is achieved, so that
\[ {\bf g}_i ({q}) = - {\bf g}_{-i} ({q})  \] 
Notice, in (\ref{tangent}) $\Delta x$ is a length scale corresponding to a representative vertex spacing of the mesh. 
It is well known that, unlike a Cartesian coordinate system
in Euclidean space, the basis tangent vectors ${\bf g}_i ({q})$ ($i = 1, 2, 3$) of a curvilinear
coordinate system are not orthonormal in general.
That is, ${\bf g}_i ({q}) \cdot {\bf g}_j ({q}) \neq \delta_{ij}$.
Hence, we should further define the corresponding metric tensor based on these basis tangent vectors,
\begin{equation} 
g_{ij}({q}) \equiv {\bf g}_i ({q}) \cdot {\bf g}_j ({q}), \;\;\; i, j = 1, 2, 3 
\label{metric}
\end{equation}
as well as the volume $J$ of the cell centered at ${\bf x}({q})$,
\begin{equation}
J({q}) \equiv ({\bf g}_1 ({q}) \times {\bf g}_2 ({q})) \cdot {\bf g}_3 ({q}) 
\label{jacobian}
\end{equation}
and we can always choose a proper handedness so that $J({q}) > 0$.
Clearly $J({q})$ is a constant in space for a uniform Bravais lattice. One can easily verify that
\begin{equation} 
g({q}) \equiv det[g_{ij}({q})] = J^2({q}) 
\label{determinant}
\end{equation}
with $det[g_{ij}({q})]$ being the determinant of the metric tensor $[g_{ij}({q})]$.  
Furthermore, we can define the co-tangent basis vectors ${\bf g}^i({q})$ ($i = 1, 2, 3$) accordingly,
\begin{eqnarray}
{\bf g}^1({q}) &\equiv& {\bf g}_2({q}) \times {\bf g}_3({q}) / J({q}) \nonumber \\
{\bf g}^2({q}) &\equiv& {\bf g}_3({q}) \times {\bf g}_1({q}) / J({q}) \nonumber \\
{\bf g}^3({q}) &\equiv& {\bf g}_1({q}) \times {\bf g}_2({q}) / J({q}) 
\label{covectors}
\end{eqnarray}
Indeed, the basis tangent vectors and the co-tangent vectors are orthonormal to each other,
\[ {\bf g}_i({q}) \cdot {\bf g}^j({q}) = \delta_i^j, \;\;\; i, j = 1, 2, 3 \]
where $\delta_i^j$ is the Kronecker delta function.
Similarly, we can define the inverse metric tensor,
\begin{equation} 
g^{ij}({q}) \equiv {\bf g}^i ({q}) \cdot {\bf g}^j ({q}), \;\;\; i, j = 1, 2, 3 
\label{inverse}
\end{equation}
Clearly the inverse metric tensor is the inverse of the metric tensor, 
$[g^{ij}({q})] = [g_{ij}({q})]^{-1}$, or 
\[ \sum_{k=1}^3 g_{ik}({q}) g^{kj}({q}) = \delta_i^j \]
and $det[g^{ij}({q})] = 1/det[g_{ij}({q})]$.

Having these basic geometric quantities defined above, we can now introduce the lattice Boltzmann velocity vectors on
a general curvilinear mesh, similar to the ones on a standard Cartesian lattice,
\begin{equation}
{\bf e}_\alpha ({q}) \equiv {c}^i_\alpha {\bf g}_i({q}) \frac{\Delta x}{\Delta t}
\label{lattice}
\end{equation}
Here and thereafter, the summation convention is used for repeated Roman indices. 
For convenience, in subsequent derivations we adopt the `lattice units' convention so that $\Delta x$ = 1 and $\Delta t = 1$. 
The constant number ${c}^i_\alpha$
is either a positive or negative integer or zero, and it is
the $i$th component value of the three dimensional coordinate array 
${c}_\alpha \equiv ({c}^1_\alpha, {c}^2_\alpha, {c}^3_\alpha )$.
For example, in the so called D3Q19 with the Greek  index $\alpha$ running from $0$ to $19$,\cite{Qian} 
\[ {c}_\alpha \in \{ (0,0,0), (\pm 1, 0, 0), (0, \pm 1, 0), (0, 0, \pm 1), (\pm 1, \pm 1, 0), (\pm 1, 0, \pm 1), (0, \pm 1, \pm 1) \} \] 
As shown in the subsequent sections, a set of
necessary moment isotropy and normalization conditions
must be satisfied in order to recover the correct full Navier-Stokes hydrodynamics.\cite{ChenShan,Shan,CGO,Molvig}  These are,
when there exists a proper set of constant weights $\{ w_\alpha; \;\; \alpha = 1, \ldots , b \}$, the set of lattice component
vectors admit moment isotropy up to the 6th order, namely
\begin{eqnarray}
& & \sum_\alpha w_\alpha = 1 \nonumber \\
& & \sum_\alpha w_\alpha {c}^i_\alpha {c}^j_\alpha = T_0 \delta^{ij} \equiv T_0 \Delta^{(2), ij} \nonumber \\
& & \sum_\alpha w_\alpha {c}^i_\alpha {c}^j_\alpha {c}^k_\alpha {c}^l_\alpha
= T_0^2 [\delta^{ij}\delta^{kl} + \delta^{ik}\delta^{jl} + \delta^{il}\delta^{jk}] \equiv T_0^2\Delta^{(4), ijkl} \nonumber \\
& & \sum_\alpha w_\alpha {c}^i_\alpha {c}^j_\alpha {c}^k_\alpha {c}^l_\alpha {c}^m_\alpha {c}^n_\alpha
\nonumber \\
& & = T_0^3 [\delta^{ij} \Delta^{(4), klmn} + \delta^{ik} \Delta^{(4), jlmn} + \delta^{il} \Delta^{(4), jkmn}
\nonumber \\
& & + \delta^{im} \Delta^{(4), jkln} + \delta^{in} \Delta^{(4), jklm}] \equiv T_0^3\Delta^{(6), ijklmn}
\label{iso}
\end{eqnarray}
where the constant $T_0$ depends on the choice of a set of lattice vectors and $\delta^{ij}$ is the Kronecker delta function.
For example, $T_0 = 1/3$ for the so called D3Q19 lattice (albeit it only satisfies (\ref{iso}) up to the 4th order).
Notice, one should not confuse the three dimensional array ${c}_\alpha$ with the lattice vector
${\bf e}_\alpha ({q})$ in (\ref{lattice}).  The former is simply an array of three constant (integer) numbers, 
while the latter is a vector (with a well defined direction and magnitude) at ${q}$ on the curvilinear mesh. 

Lastly, we define a set of specific geometric quantities below that are essential for the extended LBM model,
\begin{eqnarray}
\Theta^i_j({q} + {c}_\alpha , {q}) 
\equiv [{\bf g}_j({q} + {c}_\alpha) - {\bf g}_j({q})] \cdot {\bf g}^i({q}) \nonumber \\
i, j = 1, 2, 3; \;\;\; \alpha = 0, 1, \ldots , b
\label{gamma}
\end{eqnarray}
It is easily seen that $\Theta^i_j({q} + {c}_\alpha , {q})$ vanish if the mesh is a uniform Cartesian lattice.

Once a curvilinear mesh is specified, all the geometric quantities above are fully determined and can thus be pre-computed before starting a dynamic LBM simulation. 

\subsection{Volumetric lattice Boltzmann model on a curvilinear mesh}  

Although the basic theoretical framework of the work is more general, for simplicity of describing the basic concept,
we present the formulation for the so called isothermal LBM in this section.
Similar to the standard lattice Boltzmann equation (LBE), we write the evolution of 
particle distribution below,\cite{FHP1,SChen,Benzi,CCM,Qian}
\begin{equation}
N_\alpha({q} + {c}_\alpha, t + 1) = N_\alpha({q}, t) + \Omega_\alpha({q}, t) + \delta N_\alpha({q}, t)
\label{lbe}
\end{equation}
where $N_\alpha({q}, t)$ is the number of particles belonging to the discrete direction 
${c}_\alpha$ in the cell ${q}$ at time $t$. We have assumed in (\ref{lbe})
a unity time increment (i.e., $\Delta t = 1$)
without loss of generality. $\Omega_\alpha({q}, t)$ in (\ref{lbe}) is the collision term that
satisfies local mass and momentum conservations,
\begin{eqnarray}
\sum_\alpha \Omega_\alpha({q}, t) &=& 0 \nonumber \\
\sum_\alpha {\bf e}_\alpha ({q}) \Omega_\alpha({q}, t) &=& 0
\label{conserv}
\end{eqnarray}
The particle density distribution function $f_\alpha({q}, t)$ is related to $N_\alpha({q}, t)$, via
\begin{equation}
J({q})f_\alpha({q}, t) = N_\alpha({q}, t)
\label{dist}
\end{equation}
where $J({q})$ is the volume of cell centered at ${q}$, as defined previously. 
The fundamental fluid quantities are given by the standard hydrodynamic moments,
\begin{eqnarray}
\rho ({q}, t) &=& \sum_\alpha f_\alpha ({q}, t) \nonumber \\
\rho ({q}, t) {\bf u} ({q}, t) &=& \sum_\alpha {\bf e}_\alpha ({q}) f_\alpha ({q}, t)
\label{moments}
\end{eqnarray} 
where $\rho ({q}, t)$ and ${\bf u} ({q}, t)$ are fluid density and velocity at the location ${\bf x}({q})$ and time $t$. 
Using the relationship in (\ref{lattice}), the velocity moment above can also be rewritten as 
\begin{eqnarray}
\rho ({q}, t) {\bf u} ({q}, t) &=& \sum_\alpha {c}^i_\alpha {\bf g}_i({q}) f_\alpha ({q}, t)
\nonumber \\
&=& \rho ({q}, t) {U}^i ({q}, t) {\bf g}_i({q})
\label{velocity}
\end{eqnarray}
and the velocity component value in the curvilinear coordinate system is given by,
\begin{equation}
\rho ({q}, t) {U}^i ({q}, t) = \sum_\alpha {c}^i_\alpha f_\alpha ({q}, t)
\label{vcomp}
\end{equation}
Or for simplicity of notation, we define a three-dimensional fluid velocity array
${U}({q}, t) \equiv ({U}^1({q}, t), {U}^2({q}, t), {U}^3({q}, t))$.
Therefore, (\ref{vcomp}) can be equivalently expressed as
\begin{equation}
\rho ({q}, t) {U} ({q}, t) = \sum_\alpha {c}_\alpha f_\alpha ({q}, t)
\label{vcurvi}
\end{equation}
It is immediately seen that (\ref{vcurvi}) has the same form for the fluid velocity
as that in the standard Cartesian lattice based LBM.
Similarly, the momentum conservation of the collision term in (\ref{conserv}) can also be
written as,
\begin{equation}
\sum_\alpha {c}_\alpha \Omega_\alpha({q}, t) = 0
\label{omcons}
\end{equation}

Often in LBM the collision term takes on a linearized form,\cite{Benzi,Molvig} namely
\begin{equation}
\Omega_\alpha({q}, t) = \sum_\beta J({q}) M_{\alpha\beta} [f_\beta ({q}, t) - f^{eq}_\beta ({q}, t)]
\label{collide}
\end{equation}
where $M_{\alpha\beta}$ and $f^{eq}_\beta ({q}, t)$ represent a collision matrix 
and the equilibrium distribution function, respectively.
In particular, the so called Bhatnagar-Gross-Krook (BGK) form corresponds to 
\[ M_{\alpha\beta} = - \frac {1} {\tau} \delta_{\alpha\beta} \]
with $\tau$ being the collision relaxation time.\cite{BGK,SChen,CCM,Qian}  
In order to recover the correct Navier-Stokes hydrodynamics,
besides (\ref{conserv}) and (\ref{omcons}), the collision matrix needs to satisfy 
an additional condition,\cite{Molvig,jhon,zhang,chopad,Prandtl,Pradeep}
\begin{equation}
\sum_\alpha {c}_\alpha {c}_\alpha M_{\alpha\beta} = - \frac {1} {\tau} {c}_\beta {c}_\beta
\label{matrix}
\end{equation}
Obviously the BGK form trivially satisfies such an additional property.

The extra term $\delta N_\alpha({q}, t)$ in (\ref{lbe}) represents
the change of particle distribution due to an effective inertial force associated with the curvature and non-uniformity
of a general curvilinear mesh.  Obviously $\delta N_\alpha({q}, t)$ vanishes in
the standard LBM on a Cartesian lattice.  We explain and construct its explicit form below.

We define the advection process as an exact one-to-one hop from one site to another as in the standard
LBM, namely
\begin{equation} 
N_\alpha({q} + {c}_\alpha, t + 1) = N'_\alpha({q}, t) 
\label{adv}
\end{equation}
where $N'_\alpha({q}, t)$ is the post-collide distribution at $({q}, t)$ that is equal to the right side
of eqn.(\ref{lbe}).
Due to curvilinearity, though the amount of particles 
advected from cell ${q}$ is exactly equal to what is arrived at cell ${q} + {c}_\alpha$ (as defined in
(\ref{adv})), the momentum is in fact changed along the way. Indeed, in general
\[ {\bf e}_\alpha ({q} + {c}_\alpha) N_\alpha({q} + {c}_\alpha, t + 1) 
\neq {\bf e}_\alpha ({q}) N'_\alpha({q}, t) \]
In the above, the left side of the inequality sign is the momentum value at the end of advection process 
while the right side is the value at the beginning of the process. The inequality 
is there because the path of particles is curved (as well as stretched or compressed),
so that its velocity at the end of the advection is changed from its original value.  
This is fundamentally different from that on a uniform Cartesian lattice, in that
the particles have a constant velocity throughout the advection process.  
Consequently, we have the following inequalities in the overall momentum values,
\begin{equation}
\sum_\alpha {\bf e}_\alpha ({q}) N_\alpha({q}, t) \neq
\sum_\alpha {\bf e}_\alpha ({q}- {e}_\alpha) N'_\alpha({q} - {c}_\alpha , t - 1)
\label{momentum}
\end{equation}
where the right side of the unequal sign in (\ref{momentum}) 
represents the total amount of momentum advected
out of all the neighboring cells, while the left side
is the total momentum arrived at cell ${q}$ after the advection along the curved paths.
Thus from (\ref{momentum}) and (\ref{adv}), we see 
that the net momentum change via advection from all the neighboring cells into cell ${q}$ is given by,
\begin{equation}
J({q}) {\bm{\chi}}^I({q}, t) = - \sum_\alpha [{\bf e}_\alpha ({q}) - {\bf e}_\alpha ({q} - {c}_\alpha)]
N_\alpha({q}, t)
\label{forceI}
\end{equation}
Similarly, we can realize that the net momentum change via advection out of cell ${q}$ to 
all its neighboring cells is given by
\begin{equation}
J({q}) {\bm{\chi}}^o({q}, t) = - \sum_\alpha [{\bf e}_\alpha ({q} + {c}_\alpha) - {\bf e}_\alpha ({q})]
N'_\alpha({q}, t)
\label{forceO}
\end{equation}
Subsequently, if we impose the constraints on $\delta N_\alpha({q}, t)$ below,
\begin{eqnarray}
\sum_\alpha \delta N_\alpha({q}, t) &=& 0 \nonumber \\
\sum_\alpha {\bf e}_\alpha ({q}) \delta N_\alpha({q}, t) &=& J({q}) {\bm{\chi}}({q}, t) 
\label{constraints}
\end{eqnarray}
then we preserve an exact mass conservation as well as recover the exact momentum conservation
in discrete space-time for the underlying Euclidean space.
Here ${\bm{\chi}}({q}, t) = [{\bm{\chi}}^I({q}, t) + {\bm{\chi}}^o({q}, t)]/2$.
More specifically, the first constraint in (\ref{constraints}) means that
no mass source is introduced by $\delta N_\alpha({q}, t)$.  
On the other hand, the second constraint in (\ref{constraints}) introduces an ``inertial force''  
that equals exactly to the amount needed for achieving the momentum conservation in the underlying Euclidean space 
at any lattice site ${q}$ and time $t$.  The mechanism is in analogy with the continuum kinetic
theoretic description in a curved space.
Writing in the coordinate component form, we have
\[ {F}^i({q}, t) = {\bm{\chi}}({q}, t) \cdot {\bf g}^i ({q}) \]
Using the geometric quantities defined in the previous subsection, it can be immediately shown that
\begin{equation}
J({q}) {F}^i({q}, t) = - \frac {1} {2} \sum_\alpha {c}^j_\alpha
\{ \Theta^i_j({q} + {c}_\alpha , {q}) N'_\alpha({q}, t) 
- \Theta^i_j({q} - {c}_\alpha , {q}) N_\alpha({q}, t) \}
\label{compo-constr}
\end{equation}
where the geometric function $\Theta^i_j({q} + {c}_\alpha , {q})$ is defined in (\ref{gamma}). From (\ref{compo-constr})
and (\ref{gamma}), one can immediately see that ${F}^i({q}, t)$ vanishes if the curvilinear mesh is
a regular uniform Cartesian lattice, as is obviously the case for a basic LBM.
The second constraint in (\ref{constraints}) can also be expressed in coordinate component form as
\begin{equation}
\sum_\alpha {c}^i_\alpha \delta N_\alpha({q}, t) = J({q}) {F}^i({q}, t) 
\label{constraint2}
\end{equation}

As to be demonstrated in the next section, in order to recover the full viscous Navier-Stokes equation, an additional
constraint on the momentum flux also needs to be imposed below,
\begin{equation} 
\sum_\alpha {c}^i_\alpha {c}^j_\alpha \delta N_\alpha({q}, t) 
= J({q}) [ \delta{\Pi}^{ij}({q}, t) + \delta{\Pi}^{ji}({q}, t) ]
\label{constraint3}
\end{equation}
with
\begin{equation}
\delta{\Pi}^{ij}({q}, t) \equiv - \frac {1} {2} (1 - \frac {1} {2\tau})\sum_\alpha {c}^i_\alpha {c}^k_\alpha
[\Theta^j_k({q} + {c}_\alpha , {q}) - \Theta^j_k({q} - {c}_\alpha , {q})] 
f^{eq}_\alpha({q}, t) 
\label{delta-flux}
\end{equation}
A specific form of $\delta N_\alpha({q}, t)$ can be chosen below
\begin{equation}
\delta N_\alpha({q}, t) = w_{\alpha}J({q}) [\frac {{c}^j_\alpha {F}^j({q}, t)} {T_0}
+ (\frac {{c}^j_\alpha {c}^k_\alpha} {T_0} - \delta^{jk}) \frac {\delta{\Pi}^{jk}({q}, t)} {T_0} ]
\label{deltaN}
\end{equation}
With a simple algebra, one can verify that it satisfies 
the moment constraints of (\ref{constraints}), (\ref{constraint2}) and (\ref{constraint3}). Notice, due to the appearance of $N'_\alpha({q}, t)$ in (\ref{compo-constr}), the overall collision process for determining $N'_\alpha({q}, t)$ defines an implicit relationship, so that various ways to handle it need to be explored depending on types flows of interest. 

The final part for complete specifying the extended LBM
on a curvilinear mesh is the form of the equilibrium distribution function, which needs to be defined 
appropriately in order to recover the correct Euler equation as well as the Navier-Stokes equation in curvilinear coordinates
in the hydrodynamic limit. In particular,
the following fundamental conditions on hydrodynamic moments must be realized,
\begin{eqnarray}
& & \sum_\alpha f^{eq}_\alpha = \rho \; , \nonumber \\
& & \sum_\alpha {c}^i_\alpha f^{eq}_\alpha = \rho {U}^i \; , \nonumber \\
& & \sum_\alpha {c}^i_\alpha {c}^j_\alpha f^{eq}_\alpha \equiv {\Pi}^{ij,eq}
= g^{ij} \rho T_0 + \rho {\tilde U}^i {\tilde U}^j \; , \nonumber \\
& & \sum_\alpha {c}^i_\alpha {c}^j_\alpha {c}^k_\alpha f^{eq}_\alpha \equiv {Q}^{ijk,eq}
= [g^{ij}{\tilde U}^k + g^{jk}{\tilde U}^i + g^{ki}{\tilde U}^j]\rho T_0 + \rho {\tilde U}^i {\tilde U}^j {\tilde U}^k
\label{equilmoments}
\end{eqnarray}
In the above, 
\[ {\tilde U}^i({q}, t) = {U}^i({q}, t) + \frac {1} {2} a^i({q}, t) \]
with $\rho ({q}, t) a^i({q}, t) \equiv {F}^i({q}, t)$.  It is straightforward to show
that these fundamental conditions are met by the following equilibrium distribution form,
\begin{eqnarray}
f^{eq}_\alpha = & & \rho w_\alpha \{ 1 + \frac {{c}^i_\alpha {U}^i} {T_0}
+ \frac {1} {2T_0} (\frac {{c}^i_\alpha {c}^j_\alpha} {T_0} - \delta^{ij})
[(g^{ij} - \delta^{ij})T_0 + {\tilde U}^i {\tilde U}^j] \nonumber \\
& & + \frac {1} {6T_0^3} ({c}^i_\alpha {c}^j_\alpha {c}^k_\alpha 
- T_0 ({c}^i_\alpha \delta^{jk} + {c}^j_\alpha \delta^{ki} + {c}^k_\alpha \delta^{ij}))
[ T_0 [ (g^{ij}{\tilde U}^k - \delta^{ij}U^k) \nonumber \\
& & + (g^{jk}{\tilde U}^i - \delta^{jk}U^i) + (g^{ki}{\tilde U}^j - \delta^{ki}U^j)]
+ {\tilde U}^i {\tilde U}^j {\tilde U}^k ]  \}
\label{equilibrium}
\end{eqnarray}
The equilibrium distribution form above is analogous to that of 
a low Mach number expansion of the Maxwell-Boltzmann distribution expressed in curvilinear coordinates.\cite{veden}
It reduces to the standard LBM equilibrium distribution if the curvilinear mesh is
a uniform Cartesian lattice so that $g^{ij} = \delta^{ij}$.\cite{Molvig,Shan,ChenShan}

With all the quantities and dynamic properties  defined above, we can show that the lattice Boltzmann equation (\ref{lbe})
computed on the (non-Euclidean) uniform Cartesian lattice $\{ {q} \}$ obeys the Navier-Stokes hydrodynamics in
curvilinear coordinates.  Consequently, mapping of the resulting fluid values, $\rho ({q}, t)$ and ${U}^i({q}, t)$, 
onto the curvilinear mesh is given by a simple transformation below,
\begin{eqnarray}
\rho ({\bf x}({q}), t) &=& \rho ({q}, t) \; ,  \nonumber \\
{\bf u}({\bf x}({q}), t) &=& {U}^i({q}, t) {\bf g}_i ({q})
\label{result}
\end{eqnarray}

\section{Derivation of the Navier-Stokes Hydrodynamics in Curvilinear Coordinates}        

In this section, we provide a detailed derivation and show that the extended LBM
presented above indeed produces the correct Navier-Stokes equations in general curvilinear
coordinates.  We rewrite the lattice Boltzmann Equation (\ref{lbe}) below, 
\begin{equation}
N_\alpha({q} + {c}_\alpha, t + 1) = N_\alpha({q}, t) + \Omega_\alpha({q}, t) + \delta N_\alpha({q}, t)
\label{lbe1}
\end{equation}
Expanding it in both time and space up to the second order leads to
\begin{equation}
[\partial_t + {c}^i_\alpha \frac {\partial} {\partial q^i}
+ \frac {1} {2} (\partial_t + {c}^i_\alpha \frac {\partial} {\partial q^i})^2 ] N_\alpha 
= \Omega_\alpha + \delta N_\alpha
\label{lbe2}
\end{equation}
Now we introduce multiple scales in time and space based on the conventional Chapman-Enskog expansion procedure,\cite{CE,FHP1}
\[ \partial_t = \epsilon \partial_{t_0} + \epsilon^2 \partial_{t_1} ; \;\;\; 
\frac {\partial} {\partial q^i} = \epsilon \frac {\partial} {\partial q^i} \]
and
\[ N_\alpha = N^{eq}_\alpha + \epsilon N^{(1)}_\alpha + \epsilon^2 N^{(2)}_\alpha + \cdots \]
Here $\epsilon$ ($ << 1$) denotes a small number.
Conservation laws require,
\begin{eqnarray}
& & \sum_\alpha N^{eq}_\alpha = \sum_\alpha N_\alpha = J\rho , 
\;\;\;\; \sum_\alpha {c}^i_\alpha N^{eq}_\alpha = \sum_\alpha {c}^i_\alpha N_\alpha = J\rho {U}^i 
\nonumber \\
& & \sum_\alpha N^{(n)}_\alpha = \sum_\alpha {c}^i_\alpha N^{(n)}_\alpha = 0; \;\;\; n > 0
\label{highmoments} 
\end{eqnarray}
Likewise, $\delta N_\alpha = \epsilon \delta N^{(0)}_\alpha + \epsilon^2 \delta N^{(1)}_\alpha + \cdots $, and
\[ \sum_\alpha \delta N^{(n)}_\alpha = 0; \;\;\; n = 0, 1, \ldots \]
for they introduce no source of mass.
Equating the same powers of $\epsilon$, eqn.(\ref{lbe2}) leads to the following two equations
\begin{equation}
(\partial_{t_0} + {c}^i_\alpha \frac {\partial} {\partial q^i}) N^{eq}_\alpha 
= \sum_\beta M_{\alpha\beta} N^{(1)}_\beta + \delta N^{(0)}_\alpha
\label{zero-order}
\end{equation} 
and
\begin{eqnarray}
\partial_{t_1} N^{eq}_\alpha &+& \frac {1} {2} (\partial_{t_0} + {c}^i_\alpha \frac {\partial} {\partial q^i})^2 N^{eq}_\alpha
+ (\partial_{t_0} + {c}^i_\alpha \frac {\partial} {\partial q^i}) N^{(1)}_\alpha \nonumber \\
&=& \sum_\beta M_{\alpha\beta} N^{(2)}_\beta + \delta N^{(1)}_\alpha
\label{first-order}
\end{eqnarray}
where the collision term has taken the linearized form in (\ref{collide}).  Eqn.(\ref{zero-order}) can be directly
inverted,
\begin{equation}
N^{(1)}_\alpha = \sum_\beta M^{-1}_{\alpha\beta} [(\partial_{t_0} + {c}^i_\beta \frac {\partial} {\partial q^i}) N^{eq}_\beta 
- \delta N^{(0)}_\beta ]
\label{invert}
\end{equation}
Therefore, eqn.(\ref{first-order}) can also be written as,
\begin{eqnarray}
\partial_{t_1} N^{eq}_\alpha &+& \frac {1} {2} (\partial_{t_0} + {c}^i_\alpha \frac {\partial} {\partial q^i}) 
[ \sum_\beta M_{\alpha\beta} N^{(1)}_\beta + \delta N^{(0)}_\alpha ] \nonumber \\
&+& (\partial_{t_0} + {c}^i_\alpha \frac {\partial} {\partial q^i}) N^{(1)}_\alpha \nonumber \\
&=& \sum_\beta M_{\alpha\beta} N^{(2)}_\beta + \delta N^{(1)}_\alpha
\label{first-order2}
\end{eqnarray} 

Taking the mass moment of eqn.(\ref{zero-order}) and (\ref{first-order2}), 
and using mass conservation properties of $M_{\alpha\beta}$ and $\delta N^{(n)}$ above, we obtain
for the leading order
\begin{equation}
\sum_\alpha (\partial_{t_0} + {c}^i_\alpha \frac {\partial} {\partial q^i}) N^{eq}_\alpha = 0
\label{moment00}
\end{equation}
Based on the equilibrium moment definitions (\ref{highmoments}), eqn.(\ref{moment00}) is equivalent to
\begin{equation}
\partial_{t_0} (J\rho ) + \frac {\partial} {\partial q^i} (J \rho {U}^i) = 0
\label{moment000}
\end{equation}
For the next order, we have
\begin{equation}
\partial_{t_1} \sum_\alpha N^{eq}_\alpha
+ \frac {1} {2} \frac {\partial} {\partial q^i}\sum_\alpha {c}^i_\alpha \delta N^{(0)}_\alpha  = 0
\label{moment01}
\end{equation}
Combining (\ref{moment00}) and (\ref{moment01}), we get
\begin{equation}
\partial_t \sum_\alpha N^{eq}_\alpha + \frac {\partial} {\partial q^i} \sum_\alpha {c}^i_\alpha N^{eq}_\alpha
+ \frac {1} {2} \frac {\partial} {\partial q^i}\sum_\alpha {c}^i_\alpha \delta N^{(0)}_\alpha  = 0
\label{moment-comb}
\end{equation}
Substitute the definitions in (\ref{highmoments}), eqn.(\ref{moment-comb}) becomes
\begin{equation}
\partial_t (J\rho) + \frac {\partial} {\partial q^i} [J\rho \; ({U}^i + \frac {{a}^{(0),i}} {2} )] = 0
\label{continuity}
\end{equation}
where
\begin{equation}
\sum_\alpha {c}^i_\alpha \delta N^{(0)}_\alpha \equiv J\rho {a}^{(0),i}
\label{accele0}
\end{equation} 
is to be further discussed later in this section.  
Define the fluid velocity as 
\begin{equation}
{\tilde U}^i \equiv {U}^i + \frac {1} {2} {a}^{(0),i}
\label{utilde}
\end{equation}
and since $J = J({q})$ is not a function of time,
eqn.(\ref{continuity}) is in fact the standard mass continuity equation expressed in general
curvilinear coordinates,
\begin{equation}
\partial_t \rho + \frac {1} {J} \frac {\partial} {\partial q^i} (J\rho {\tilde U}^i ) = 0
\label{continuity1}
\end{equation}
Recognizing that $\frac {1} {J} \frac {\partial} {\partial q^i} (J A^i) = \nabla \cdot {\bf A}$ is the
divergence of a vector ${\bf A}$ in a general curvilinear coordinate system, eqn.(\ref{continuity1}) is simply
the familiar mass continuity equation in the coordinate free representation,
\begin{equation}
\partial_t \rho + \nabla \cdot (\rho {\tilde {\bf u}}) = 0
\label{continuity12}
\end{equation}

Taking the momentum moment of eqn.(\ref{zero-order}), we get,
\begin{equation}
\sum_\alpha {c}^i_\alpha (\partial_{t_0} + {c}^j_\alpha \frac {\partial} {\partial q^j}) N^{eq}_\alpha 
= \sum_\alpha {c}^i_\alpha \delta N^{(0)}_\alpha
\label{mconti}
\end{equation}
Define (see also in (\ref{equilmoments}))
\begin{equation}
J{\Pi}^{ij,eq} \equiv \sum_\alpha {c}^i_\alpha {c}^j_\alpha N^{eq}_\alpha
\label{pi0}
\end{equation}
together with (\ref{accele0}), eqn.(\ref{mconti}) becomes
\begin{equation}
\partial_{t_0} (J\rho {U}^i ) + \frac {\partial} {\partial q^j} (J {\Pi}^{ij,eq}) 
= J\rho {a}^{(0),i}
\label{mconti2}
\end{equation}
Similarly, taking the momentum moment of eqn.(\ref{first-order2}), 
\begin{eqnarray}
\partial_{t_1} \sum_\alpha {c}^i_\alpha N^{eq}_\alpha 
&+& \sum_\alpha {c}^i_\alpha {c}^j_\alpha \frac {\partial} {\partial q^j} 
\sum_\beta [\delta_{\alpha\beta} + \frac {1} {2} M_{\alpha\beta} ] N^{(1)}_\beta \nonumber \\
&+& \frac {1} {2} \sum_\alpha {c}^i_\alpha (\partial_{t_0} + {c}^j_\alpha \frac {\partial} {\partial q^j}) \delta N^{(0)}_\alpha 
\nonumber \\
&=& \sum_\alpha {c}^i_\alpha \delta N^{(1)}_\alpha
\label{mconti3}
\end{eqnarray} 
Let 
\begin{equation}
J{\Pi}^{ij,(1)} \equiv \sum_\alpha {c}^i_\alpha {c}^j_\alpha \sum_\beta
(\delta_{\alpha\beta} + \frac {1} {2} M_{\alpha\beta} ) N^{(1)}_\beta
\label{pi1}
\end{equation}
and use relation (\ref{invert}), we can derive
\begin{eqnarray}
J{\Pi}^{ij,(1)} &=& \sum_\alpha {c}^i_\alpha {c}^j_\alpha 
\sum_\beta (\delta_{\alpha\beta} + \frac {1} {2} M_{\alpha\beta} )
\sum_\gamma M^{-1}_{\beta\gamma} [(\partial_{t_0} + {c}^k_\gamma \frac {\partial} {\partial q^k}) N^{eq}_\gamma
- \delta N^{(0)}_\gamma ] \nonumber \\
&=& \sum_\alpha {c}^i_\alpha {c}^j_\alpha 
\sum_\beta (M^{-1}_{\alpha\beta} + \frac {1} {2} \delta_{\alpha\beta} )
[(\partial_{t_0} + {c}^k_\beta \frac {\partial} {\partial q^k}) N^{eq}_\beta
- \delta N^{(0)}_\beta ] 
\label{pi1-1}
\end{eqnarray}
Using the collision matrix property (\ref{matrix}) and its inverse, namely
\begin{equation}
\sum_\alpha {c}_\alpha {c}_\alpha M_{\alpha\beta} = - \frac {1} {\tau} {c}_\beta {c}_\beta; \;\;\;\;
\sum_\alpha {c}_\alpha {c}_\alpha M^{-1}_{\alpha\beta} = - \tau {c}_\beta {c}_\beta
\label{matrix2}
\end{equation}
then eqn.(\ref{pi1-1}) becomes
\begin{equation}
J{\Pi}^{ij,(1)} = - (\tau - \frac {1} {2}) \sum_\alpha {c}^i_\alpha {c}^j_\alpha
[(\partial_{t_0} + {c}^k_\alpha \frac {\partial} {\partial q^k}) N^{eq}_\alpha - \delta N^{(0)}_\alpha ] 
\label{pi1-2}
\end{equation}
Substitute $J{\Pi}^{ij,(1)}$ of the above into (\ref{mconti3}), we get
\begin{eqnarray}
\partial_{t_1} \sum_\alpha {c}^i_\alpha N^{eq}_\alpha 
&-& \frac {\partial} {\partial q^j} \{ (\tau - \frac {1} {2})\sum_\alpha {c}^i_\alpha {c}^j_\alpha 
[(\partial_{t_0} + {c}^k_\alpha \frac {\partial} {\partial q^k}) N^{eq}_\alpha - \delta N^{(0)}_\alpha ]\} \nonumber \\
&+& \frac {1} {2} \sum_\alpha {c}^i_\alpha (\partial_{t_0} + {c}^j_\alpha \frac {\partial} {\partial q^j}) \delta N^{(0)}_\alpha 
\nonumber \\
&=& \sum_\alpha {c}^i_\alpha \delta N^{(1)}_\alpha
\label{mconti4}
\end{eqnarray} 
After some simple cancellations and rearrangements, it becomes
\begin{eqnarray}
\partial_{t_1} \sum_\alpha {c}^i_\alpha N^{eq}_\alpha 
&+& \frac {1} {2} \partial_{t_0} \sum_\alpha {c}^i_\alpha \delta N^{(0)}_\alpha \nonumber \\
&-& \frac {\partial} {\partial q^j} [ (\tau - \frac {1} {2})\sum_\alpha {c}^i_\alpha {c}^j_\alpha 
(\partial_{t_0} + {c}^k_\alpha \frac {\partial} {\partial q^k}) N^{eq}_\alpha ] \nonumber \\
&+& \frac {\partial} {\partial q^j} [\tau \sum_\alpha {c}^i_\alpha {c}^j_\alpha \delta N^{(0)}_\alpha ]
\nonumber \\
&=& \sum_\alpha {c}^i_\alpha \delta N^{(1)}_\alpha
\label{mconti5}
\end{eqnarray} 
Using definitions in (\ref{equilmoments}), $N^{eq}_\alpha = Jf^{eq}_\alpha$, and (\ref{accele0}), eqn.(\ref{mconti5})
can be equivalently expressed as
\begin{eqnarray}
\partial_{t_1} (J\rho {U}^i ) 
&+& \frac {1} {2} \partial_{t_0} (J\rho {a}^{(0),i}) \nonumber \\
&-& \frac {\partial} {\partial q^j} [ (\tau - \frac {1} {2})\sum_\alpha {c}^i_\alpha {c}^j_\alpha 
(\partial_{t_0} + {c}^k_\alpha \frac {\partial} {\partial q^k}) N^{eq}_\alpha ] \nonumber \\
&+& \frac {\partial} {\partial q^j} [\tau \sum_\alpha {c}^i_\alpha {c}^j_\alpha \delta N^{(0)}_\alpha ]
\nonumber \\
&=& J\rho {a}^{(1),i}
\label{mconti6}
\end{eqnarray} 
where in the above we have defined $J\rho {a}^{(1),i} \equiv \sum_\alpha {c}^i_\alpha \delta N^{(1)}_\alpha$.

Now let us examine the forcing terms $J\rho {a}^{(0),i}$ and $J\rho {a}^{(1),i}$. Recall from the momentum
conservation condition, the overall forcing term in eqn.(\ref{lbe}) is given by (\ref{constraint2}), that is
\[ \sum_\alpha {c}^i_\alpha \delta N_\alpha({q}, t) = J({q}) {F}^i({q}, t) \]
with $J{F}^i$ given by (\ref{compo-constr}),
\begin{equation}
J({q}) {F}^i({q}, t) = - \frac {1} {2} \sum_\alpha {c}^j_\alpha
\{ \Theta^i_j({q} + {c}_\alpha , {q}) N'_\alpha({q}, t) 
- \Theta^i_j({q} - {c}_\alpha , {q}) N_\alpha({q}, t) \}
\label{compo-constrA}
\end{equation}
where $\Theta^i_j({q} + {c}_\alpha , {q})$ is defined in (\ref{gamma}). 
Taking the continuum limit, we have for the leading order of $\Theta^i_j({q} + {c}_\alpha , {q})$ 
\begin{eqnarray}
\Theta^i_j({q} + {c}_\alpha , {q}) 
&\equiv& [{\bf g}_j({q} + {c}_\alpha) - {\bf g}_j({q})] \cdot {\bf g}^i({q}) \nonumber \\
&\approx& {c}^k_\alpha\frac {\partial {\bf g}_j({q})} {\partial q^k} \cdot {\bf g}^i({q})
\label{gammaA}
\end{eqnarray}
Similarly, 
\begin{eqnarray}
\Theta^i_j({q} - {c}_\alpha , {q}) 
&\equiv& [{\bf g}_j({q} - {c}_\alpha) - {\bf g}_j({q})] \cdot {\bf g}^i({q}) \nonumber \\
&\approx& - {c}^k_\alpha\frac {\partial {\bf g}_j({q})} {\partial q^k} \cdot {\bf g}^i({q})
\label{gammaB}
\end{eqnarray}
Consequently, when substitute (\ref{gammaA}) and (\ref{gammaB}) in (\ref{compo-constrA}), we obtain, to the leading order
\begin{equation}
J({q}) {F}^i({q}, t) = - \frac {1} {2} \Gamma^i_{jk} ({q}) \sum_\alpha {c}^j_\alpha {c}^k_\alpha
[N'_\alpha({q}, t) + N_\alpha({q}, t) ]
\label{compo-constrB}
\end{equation}
where
\begin{equation}
\Gamma^i_{jk} ({q}) \equiv \frac {\partial {\bf g}_j({q})} {\partial q^k} \cdot {\bf g}^i({q})
\label{Christoffle}
\end{equation}
is the Christoffel symbol as defined in differential geometry.\cite{book}
Also in the above, the relationship (\ref{adv}) is used, namely
\begin{equation} 
N_\alpha({q} + {c}_\alpha , t + 1) = N'_\alpha({q}, t) 
\label{advA}
\end{equation}
Furthermore, since 
\[ N_\alpha({q} + {c}_\alpha , t + 1) \approx N_\alpha({q}, t) + 
(\partial_{t_0} + {c}^k_\alpha \frac {\partial} {\partial q^k}) N_\alpha({q}, t) \]
we have from (\ref{compo-constrB}) the following
\begin{eqnarray}
\sum_\alpha {c}^i_\alpha \delta N_\alpha({q}, t) &=& J({q}) {F}^i({q}, t) \nonumber \\
&=& - \Gamma^i_{jk} ({q}) \sum_\alpha {c}^j_\alpha {c}^k_\alpha N_\alpha({q}, t)
\nonumber \\
&-& \frac {1} {2} \Gamma^i_{jk} ({q}) \sum_\alpha {c}^j_\alpha {c}^k_\alpha
(\partial_{t_0} + {c}^l_\alpha \frac {\partial} {\partial q^l}) N_\alpha({q}, t) 
\label{compo-constrC}
\end{eqnarray}
Comparing terms of the same order in $\epsilon$, we recognize that
\begin{equation}
\sum_\alpha {c}^i_\alpha \delta N^{(0)}_\alpha = J\rho {a}^{(0),i} 
= - \Gamma^i_{jk} \sum_\alpha {c}^j_\alpha {c}^k_\alpha N^{eq}_\alpha
\label{a0}
\end{equation}
and
\begin{eqnarray}
\sum_\alpha {c}^i_\alpha \delta N^{(1)}_\alpha &=& J\rho {a}^{(1),i} \nonumber \\ 
&=& - \Gamma^i_{jk} \sum_\alpha {c}^j_\alpha {c}^k_\alpha N^{(1)}_\alpha \\
&-& \frac {1} {2} \Gamma^i_{jk} \sum_\alpha {c}^j_\alpha {c}^k_\alpha
(\partial_{t_0} + {c}^l_\alpha \frac {\partial} {\partial q^l}) N^{eq}_\alpha
\label{a1}
\end{eqnarray}
Replacing $N^{(1)}_\alpha$ with that in (\ref{invert}), with the use of property (\ref{matrix2})
and after some straightforward algebra, eqn.(\ref{a1}) becomes
\begin{eqnarray}
\sum_\alpha {c}^i_\alpha \delta N^{(1)}_\alpha &=& J\rho {a}^{(1),i} \nonumber \\ 
&=& \Gamma^i_{jk} \{ (\tau - \frac {1} {2}) \sum_\alpha {c}^j_\alpha {c}^k_\alpha 
(\partial_{t_0} + {c}^l_\alpha \frac {\partial} {\partial q^l}) N^{eq}_\alpha
- \tau \sum_\alpha {c}^j_\alpha {c}^k_\alpha \delta N^{(0)}_\alpha \}
\label{a11}
\end{eqnarray}

From the definition (\ref{pi0}) and (\ref{a0}), we see that 
\begin{equation}
J\rho {a}^{(0),i} = - \Gamma^i_{jk} J{\Pi}^{jk,eq} 
\label{gamma-pi0}
\end{equation}
Similarly, let us also define
\begin{eqnarray}
J{\Pi}^{ij,neq} &\equiv& - (\tau - \frac {1} {2}) \sum_\alpha {c}^i_\alpha {c}^j_\alpha 
(\partial_{t_0} + {c}^k_\alpha \frac {\partial} {\partial q^k}) N^{eq}_\alpha \nonumber \\
&+& \tau \sum_\alpha {c}^i_\alpha {c}^j_\alpha \delta N^{(0)}_\alpha \nonumber \\
&=& - (\tau - \frac {1} {2})  
[\partial_{t_0} (J{\Pi}^{ij,eq}) + \frac {\partial} {\partial q^k} (J{Q}^{ijk,eq})] \nonumber \\
&+& \tau \sum_\alpha {c}^i_\alpha {c}^j_\alpha \delta N^{(0)}_\alpha
\label{pneq}
\end{eqnarray}
where
\begin{equation}
J{Q}^{ijk,eq} \equiv \sum_\alpha {c}^i_\alpha {c}^j_\alpha {c}^k_\alpha N^{eq}_\alpha
\label{heatf}
\end{equation}
Hence, from (\ref{a11}), we have
\begin{equation}
J\rho {a}^{(1),i} = - \Gamma^i_{jk} J{\Pi}^{jk,neq} 
\label{gamma-pneq}
\end{equation}

Taking (\ref{gamma-pi0}) to eqn.(\ref{mconti2}), we have for the leading order the equation below
\begin{equation}
\partial_{t_0} (J\rho {U}^i ) + \frac {\partial} {\partial q^j} (J{\Pi}^{ij,eq}) + \Gamma^i_{jk} J{\Pi}^{jk,eq}
= 0
\label{euler}
\end{equation}
On the other hand, taking (\ref{pneq}) and (\ref{gamma-pneq}) to eqn.(\ref{mconti6}), we get in the viscous order
\begin{equation}
\partial_{t_1} (J\rho {U}^i ) 
+ \frac {1} {2} \partial_{t_0} (J\rho {a}^{(0),i}) 
+ \frac {\partial} {\partial q^j} (J{\Pi}^{ij,neq}) + \Gamma^i_{jk} J{\Pi}^{jk,neq}
= 0
\label{viscous}
\end{equation}
Combining (\ref{euler}) and (\ref{viscous}) together and using the definition in (\ref{utilde}),
one obtains formally the Cauchy's transport equation, namely
\begin{equation}
\partial_t (\rho {\tilde U}^i ) 
+ \frac{1} {J} \frac {\partial} {\partial q^j} (J{\Pi}^{ij}) + \Gamma^i_{jk} {\Pi}^{jk} = 0
\label{chauchy}
\end{equation}
where ${\Pi}^{jk} = {\Pi}^{jk,eq} + {\Pi}^{jk,neq}$.

Equation (\ref{chauchy}) can be further expressed in terms of hydrodynamic quantities.  From the 
moment properties of the equilibrium distribution function in (\ref{equilmoments}), we have
\begin{equation}
{\Pi}^{ij,eq} = g^{ij} \rho T_0  + \rho {\tilde U}^i {\tilde U}^j 
\label{hydro-peq}
\end{equation}
which has exactly the same form as that from the continuum kinetic theory.
On the other hand, from (\ref{pneq}) we have
\begin{eqnarray}
{\Pi}^{ij,neq} &=& - (\tau - \frac {1} {2})  
[\partial_{t_0} {\Pi}^{ij,eq} + \frac {1} {J} \frac {\partial} {\partial q^k} (J{Q}^{ijk,eq})] \nonumber \\
&+& \frac{\tau}{J} \sum_\alpha {c}^i_\alpha {c}^j_\alpha \delta N^{(0)}_\alpha 
\label{hydro-pneq}
\end{eqnarray}
where, from (\ref{equilmoments}), ${\Pi}^{ij,eq}$ is 
given by (\ref{hydro-peq}) and ${Q}^{ijk,eq}$ is given below
\begin{equation}
{Q}^{ijk,eq} 
= [g^{ij}{\tilde U}^k + g^{jk}{\tilde U}^i + g^{ki}{\tilde U}^j]\rho T_0 + \rho {\tilde U}^i {\tilde U}^j {\tilde U}^k  
\label{hydro-qeq}
\end{equation}
which has exactly the same form as that from the continuum kinetic theory.
However, the term $\tau \sum_\alpha {c}^i_\alpha {c}^j_\alpha \delta N^{(0)}_\alpha$
needs to be further evaluated in terms of the hydrodynamic quantities.

From (\ref{constraint3}) and (\ref{delta-flux}), we have
\begin{equation} 
\sum_\alpha {c}^i_\alpha {c}^j_\alpha \delta N_\alpha({q}, t) 
= J({q}) [ \delta{\Pi}^{ij}({q}, t) + \delta{\Pi}^{ji}({q}, t) ]
\label{constraint3a}
\end{equation}
with
\begin{equation}
\delta{\Pi}^{ij}({q}, t) \equiv - \frac {1} {2} (1 - \frac {1} {2\tau})\sum_\alpha {c}^j_\alpha {c}^k_\alpha
[\Theta^i_k({q} + {c}_\alpha , {q}) - \Theta^i_k({q} - {c}_\alpha , {q})] 
f^{eq}_\alpha({q}, t) 
\label{delta-fluxa}
\end{equation}
Taking the long wave length limit, (\ref{delta-fluxa}) in the continuum limit becomes
\begin{eqnarray}
\delta{\Pi}^{ij}({q}, t) &\approx& - (1 - \frac {1} {2\tau})\Gamma^i_{kl}({q})
\sum_\alpha {c}^j_\alpha {c}^k_\alpha {c}^l_\alpha f^{eq}_\alpha({q}, t) \nonumber \\
&=& - (1 - \frac {1} {2\tau})\Gamma^i_{kl}({q}) {Q}^{jkl,eq}({q}, t) 
\label{delta-fluxb}
\end{eqnarray}
Therefore, 
\begin{equation}
\tau \sum_\alpha {c}^i_\alpha {c}^j_\alpha \delta N^{(0)}_\alpha
= - (\tau - \frac {1} {2}) J({q}) [\Gamma^i_{kl}({q}) {Q}^{jkl,eq}({q}, t) 
+ \Gamma^j_{kl}({q}) {Q}^{ikl,eq}({q}, t) ]
\label{ddN}
\end{equation}
So we finally obtain from (\ref{hydro-pneq}) that
\begin{eqnarray}
{\Pi}^{ij,neq} &=& - (\tau - \frac {1} {2}) \{
\partial_{t_0} {\Pi}^{ij,eq} + \frac {1} {J} \frac {\partial} {\partial q^k} (J{Q}^{ijk,eq}) \nonumber \\
&+& [\Gamma^i_{kl}({q}) {Q}^{jkl,eq} + \Gamma^j_{kl}({q}) {Q}^{ikl,eq} ] \}
\label{hydro-pneq1}
\end{eqnarray}
with ${\Pi}^{ij,eq}$ and ${Q}^{ijk,eq}$ given by (\ref{hydro-peq}) and (\ref{hydro-qeq}).

Substituting eqns.(\ref{hydro-peq}) and (\ref{hydro-pneq1}) into eqn.(\ref{chauchy}), comparing with
eqn.(\ref{flux-lbm}) in Appendix B, together with the definitions of (\ref{hydro-peq}) and (\ref{hydro-qeq})
(having the same forms as that from the continuum kinetic theory), we finally arrive
at exactly the same form as derived out of a continuum kinetic theory in curvilinear coordinates. 
The only difference between the two is that the collision time $\tau$ is replaced by $(\tau - \frac {1} {2})$ here.
The rest of derivation for obtaining the Navier-Stokes hydrodynamics is a straightforward algebra and is no different
from that of the continuum kinetic theory in curvilinear coordinates (provided in Appendices A and B).  
Notice, same as the derivation in the continuum kinetic theory, 
in (\ref{hydro-pneq1}) the $0$th order time derivative term $\partial_{t_0} {\Pi}^{ij,eq}$ is to be replaced
by the leading (Euler) order equations on the right hand sides of eqns.(\ref{euler}) and (\ref{moment000}).  
Without repeating the rest of algebra, 
here we present the final form - the Navier-Stokes equation in a general curvilinear coordinate system,
\begin{eqnarray}
\partial_t (\rho {\tilde U}^i) &+& \frac{1} {J} \frac {\partial} {\partial q^j} (J\rho {\tilde U}^i {\tilde U}^j)
+ \Gamma^i_{jk}\rho {\tilde U}^j {\tilde U}^k \nonumber \\
&=& - g^{ij} \frac {\partial p} {\partial q^j} + 
\frac{1} {J} \frac {\partial} {\partial q^j} (2J\mu S^{ij}) + 2\Gamma^i_{jk}\mu S^{jk}
\label{NS}
\end{eqnarray}
where the pressure $p = \rho T_0$, $\mu = (\tau - \frac {1} {2})\rho T_0$, and
\[ S^{ij} = \frac {1} {2} [{\tilde U}^i|_k g^{kj} + {\tilde U}^j|_k g^{ki} ] \]
with the standard covariant derivative of the velocity component defined as
\[{\tilde U}^i|_k \equiv \frac {\partial {\tilde U}^i} {\partial q^k} + \Gamma^i_{kl}{\tilde U}^l \]
Using the standard definitions of differential operators in differential geometry,\cite{book}
we can recognize that eqn.(\ref{NS}) is indeed the familiar Navier-Stokes equation in
a generic coordinate-free operator representation, 
\begin{equation}
\partial_t (\rho {\tilde {\bf u}}) + \nabla \cdot (\rho {\tilde {\bf u}}{\tilde {\bf u}})
= - \nabla p + \nabla \cdot (2\mu {\bf S}) 
\label{NSv}
\end{equation}

\section{Discussion}

In this paper, we present a theoretical formulation of lattice Boltzmann models
in a general curvilinear  coordinate system.  The formulation is an application of the Riemannian geometry 
for kinetic theory\cite{veden}
to discrete space and time. Unlike some previous works,\cite{Mendosa1,Mendosa2,Mendosa3} 
here we use a volumetric representation so that
conservation laws are exactly ensured.\cite{volumetric} Furthermore, in the current formulation, 
we find that the main and the only additional source term in the extended LBM model is
corresponding to the inertial force to ensure the exact momentum conservation in the underlying
Euclidean space. This is the same as in the continuum kinetic theory.  
On the other hand, this forcing term needs to be applied at an appropriate discrete time in order to
realize the correct viscous fluid effect associated with non-equilibrium physics.  
The equilibrium distribution function also needs to be properly modified 
that is directly analogous to the Maxwell-Boltzmann distribution on a curved space.  
Unlike the previous formulations,\cite{Mendosa1,Mendosa2,Mendosa3} 
there are no other terms or treatments added to cancel any discrete artifacts.
Through a detailed analysis, we
have shown that the current LBM formulation recovers the correct Navier-Stokes  behavior 
in the hydrodynamic limit, as long as a discrete lattice velocity set satisfying a sufficient order of isotropy is used. Extensive numerical validations of this extended LBM for various flows on various curvilinear meshes are to be presented in future publications.
The main benefit of this kind of theoretical formulation in a general curvilinear
coordinate system is its preservation of the key LBM one-to-one Lagrangian nature of particle advection. 
Not only this is desirable in algorithmic simplicity, as a standard LBM on a Cartesian lattice, 
it also has non-trivial implications for flows at finite Knudsen number.\cite{Shan,jhon,zhang} 
Although the specific LBM model constructed here is for an isothermal fluid, 
the fundamental framework is directly extendable to more general fluid flow situations.
Possible extensions in the future may include transport of scalars, complex fluids, finite Knudsen flows, 
as well as higher speed flows with energy and temperature dynamics in curved space. 
The latter is essential for study of highly compressible flows and flows with substantial temperature variations. 
Another interesting possible extension of the current
theoretical formulation in the future is for a time varying coordinate system. This is useful particularly for studying of fluid flows around a dynamically deforming solid object.

\vspace{0.1in}

{\bf Acknowledgment:} I am grateful to Raoyang Zhang, Pradeep Gopalakrishnan, Ilya Staroselsky and Alexei Chekhlov for insightful discussions.

\appendix

\section{Boltzmann Kinetic Equation in Curvilinear Coordinates and Curved Spaces}

In the absence of an external force, a particle has a constant velocity and moves along a straight line in Euclidean
space according to the 1st law of Newton.  That is, time derivatives of the particle position and
the velocity vectors are described by
\begin{equation}
{\dot {\bf x}} = {\bf v}, \;\;\; {\dot {\bf v}} = 0
\label{firstlaw}
\end{equation}
where ${\bf x}$ and ${\bf v}$ are the position and velocity vectors, respectively.  If we express the velocity
vector in a curvilinear coordinate system, we have
\begin{equation}
{\bf v} = v^i {\bf g}_i({q})
\label{velo}
\end{equation}
Therefore,
\begin{equation}
{\dot {\bf v}} = 0 \; \rightarrow \; {\dot v^i}{\bf g}_i + v^i \frac {\partial {\bf g}_i} {\partial q^j} v^j = 0 
\label{dotv}
\end{equation}
where in the above we have used the definition $v^j = {\dot q^j}$.

Since 
\begin{equation}
\frac {\partial {\bf g}_i} {\partial q^j} = \frac {\partial {\bf g}_i} {\partial q^j} \cdot {\bf g}^k {\bf g}_k
\equiv \Gamma^k_{ij}{\bf g}_k ,
\label{gderiv}
\end{equation}
we have, by rearranging dummy indices
\begin{equation}
{\dot {\bf v}} = 0 \; \rightarrow \; {\dot v^i}{\bf g}_i + v^j v^k \Gamma^i_{jk}{\bf g}_i = 0 
\label{dotv1}
\end{equation}
Therefore, there is an effective acceleration (inertial force) in the space of coordinates, namely
\begin{equation}
{\dot v^i} = - v^j v^k \Gamma^i_{jk}
\label{vdott}
\end{equation}

Based on the properties above, we are ready to write the Boltzmann equation in curvilinear coordinates, 
\begin{equation}
\partial_t N + \frac {\partial} {\partial q^i} (v^i N) + \frac {\partial} {\partial v^i} ({\dot v^i}N) = \Omega
\label{Boltzv}
\end{equation}
where $N \equiv N({q}, {\bar v}, t)$ denotes the number of particles inside a small pocket 
($(x^i, x^i + dx^i), (v^i, v^i + dv^i); i = 1, 2, 3$) of fluid of volume $J({q})$. 
$\Omega = \Omega ({q}, {\bar v}, t)$ is the collision term as discussed in the text,\cite{veden} satisfying local mass and
momentum conservaton laws
\begin{equation}
\int d{\bar v} \Omega ({q}, {\bar v}, t) = 0, \;\;\; \int d{\bar v} v^i \Omega ({q}, {\bar v}, t) = 0, 
\label{aconserv}
\end{equation}
where the integral operator above is defined as $\int d{\bar v} \equiv \int dv^1 dv^2 dv^3$.
Substituting (\ref{vdott}) into
(\ref{Boltzv}), we get
\begin{equation}
\partial_t N + \frac {\partial} {\partial q^i} (v^i N) - \frac {\partial} {\partial v^i} (v^j v^k \Gamma^i_{jk} N) = \Omega
\label{Boltzvv}
\end{equation}

Define a particle density function
\begin{equation}
J({q}) f({q}, {\bar v}, t) \equiv N({q}, {\bar v}, t)
\label{adensity}
\end{equation}
then we have the hydrodynamic moments specified below
\begin{eqnarray}
\int d{\bar v} N({q}, {\bar v}, t) &=& J({q}) \int d{\bar v} f({q}, {\bar v}, t) = J({q}) \rho ({q}, t) 
\nonumber \\ 
\int d{\bar v} v^i N({q}, {\bar v}, t) &=& J({q}) \int d{\bar v} v^i f({q}, {\bar v}, t) 
= J({q}) \rho ({q}, t) u^i ({q}, t) 
\label{amomnts}
\end{eqnarray}

Taking the moment integral of the Boltzmann equation (\ref{Boltzvv}), 
and using the collision properties of (\ref{aconserv}), we
obtain the two continuity equations, corresponding to mass and momentum conservations respectively
\begin{eqnarray}
& & \int d{\bar v} \{ \partial_t N + \frac {\partial} {\partial q^i} (v^i N) - \frac {\partial} {\partial v^i} (v^j v^k \Gamma^i_{jk} N) \} = 0
\nonumber \\
& & \int d{\bar v} v^i \{ \partial_t N + \frac {\partial} {\partial q^j} (v^j N) - \frac {\partial} {\partial v^l} (v^j v^k \Gamma^l_{jk} N) \} = 0
\label{acontinuities}
\end{eqnarray}
The term $\int d{\bar v} \frac {\partial} {\partial v^i} (v^j v^k \Gamma^i_{jk} N) = 0$ via integration by parts, and using definition
in (\ref{amomnts}) we get the mass continuity equation as follows
\begin{equation}
\partial_t (J\rho ) + \frac {\partial} {\partial q^i} (J \rho u^i ) = 0
\label{amassconti}
\end{equation}
or in the more familiar form
\begin{equation}
\partial_t \rho + \frac {1} {J} \frac {\partial} {\partial q^i} (J \rho u^i ) = 0
\label{amassconti}
\end{equation}
since the volume $J$ is not dependent on time $t$.

Integrate by parts, and use $\frac {\partial v^i} {\partial v^j} = \delta^i_j$, 
\[ \int d{\bar v} v^i \frac {\partial} {\partial v^l} (v^j v^k \Gamma^l_{jk} N) 
= - \int d{\bar v} v^j v^k \Gamma^i_{jk} N \] 
Hence, the second equation in (\ref{acontinuities}) becomes
\begin{equation}
\partial_t (\rho u^i) + \frac {1} {J} \frac {\partial} {\partial q^j} (J \Pi^{ij} ) + \Gamma^i_{jk} \Pi^{jk} = 0
\label{amomconti}
\end{equation}
where the momentum flux tensor $\Pi^{ij} = \Pi^{ij}({q}, t)$ is defined by
\begin{equation}
\Pi^{ij} \equiv \int d{\bar v} v^iv^j f({q}, {\bar v}, t)
\label{amflux}
\end{equation}
Eqn.(\ref{amomconti}) is known as the Cauchy's transport equation. 

We can separate the momentum flux tensor into two parts associated, respectively, to the equilibrium and the non-equilibrium parts of
the distributions $f({q}, {\bar v}, t) = f^{eq}({q}, {\bar v}, t) + f^{neq}({q}, {\bar v}, t)$, so that
$\Pi^{ij}({q}, t) = \Pi^{ij, eq}({q}, t) + \Pi^{ij, neq}({q}, t)$. The equilibrium distribution is
given by the Maxwell-Boltzmann form,
\begin{equation}
f^{eq} = \rho W\; exp[- \frac {{\bf U}^2} {2\theta} ]
\label{MaxBoltz1}
\end{equation}
where ${\bf U} \equiv {\bf v} - {\bf u}$ and $\theta$ is the temperature.  In terms of curvilinear coordinates, 
\[ {\bf U}^2 = {\bf U}\cdot {\bf U} = U^i{\bf g}_i \cdot U^j{\bf g}_j = U^i g_{ij} U^j \]
Hence, we can rewrite (\ref{MaxBoltz1}) in terms of curvilinear coordinates below
\begin{equation}
f^{eq} = \rho W\; exp[- \frac {g_{ij}U^iU^j} {2\theta} ]
\label{MaxBoltz}
\end{equation}
and the normalization factor $W = 1/\sqrt{(2\theta \pi )^3 det[g^{ij}]}$.  Here, the inverse metric tensor $[g^{ij}]$
is defined such that $g^{ik} g_{kj} = \delta^i_j$ in differential geometry. $det[g^{ij}]$ is the determinant of $[g^{ij}]$.
Using a  few basic properties of Gaussian integral,
\begin{eqnarray} 
& & \int d{U} W\; exp[- \frac {g_{ij}U^iU^j} {2\theta} ] = 1 \nonumber \\
& & \int d{U} W\; U^kU^lexp[- \frac {g_{ij}U^iU^j} {2\theta} ] = g^{kl}\theta \nonumber \\
& & \int d{U} W\; U^{k_1}U^{k_2} \cdots U^{k_n} exp[- \frac {g_{ij}U^iU^j} {2\theta} ] = 0, \;\;\; n = odd \; number
\label{Gaussian}
\end{eqnarray}
we immediately obtain from (\ref{amflux}) and (\ref{MaxBoltz}) that
\begin{eqnarray}
& & \rho = \int d{\bar v} f^{eq}, \;\;\; \rho u^i = \int d{\bar v} v^i f^{eq},   
\nonumber \\
& & \Pi^{ij, eq} = \int d{\bar v} v^iv^j f^{eq} = g^{ij} \rho\theta + \rho u^iu^j
\label{EoS}
\end{eqnarray}
Therefore, at the Euler order, in which the momentum flux tensor only includes the equilibrium contribution,
eqn.(\ref{amomconti}) reduces to
\begin{equation}
\partial_t (\rho u^i) + \frac {1} {J} \frac {\partial} {\partial q^j} (J \Pi^{ij,eq} ) + \Gamma^i_{jk} \Pi^{jk,eq} = 0
\label{aEuler}
\end{equation}
More explicitly,
\begin{equation}
\partial_t (\rho u^i) + \frac {1} {J} \frac {\partial} {\partial q^j} (J [g^{ij} \rho\theta + \rho u^iu^j]) 
+ \Gamma^i_{jk} [g^{jk} \rho\theta + \rho u^ju^k] = 0
\label{aEuler1}
\end{equation}
However, since the underlying space is ``flat'' (Euclidean), the metric tensor obeys the following property
\begin{equation}
\frac {1} {J} \frac {\partial} {\partial q^j} (J g^{ij}) + \Gamma^i_{jk} g^{jk} = 0
\label{flat}
\end{equation}
Substituting (\ref{flat}) into eqn.(\ref{aEuler1}), we arrive at a more standard form of the Euler equation,
\begin{equation}
\partial_t (\rho u^i) + \frac {1} {J} \frac {\partial} {\partial q^j} (J \rho u^iu^j) 
+ \Gamma^i_{jk} \rho u^ju^k = - g^{ij} \frac {\partial p} {\partial q^j}
\label{aEuler2}
\end{equation}
with the pressure defined by an ideal gas equation of state, $p = \rho\theta$.

\section{Derivation of the Navier-Stokes Equations in General Coordinates}

To derive the Navier-Stokes hydrodynamics up to the viscous order, we use the Chapman-Enskog expansion procedure,\cite{CE}
\[ \partial_t = \epsilon \partial_{t_0} + \epsilon^2 \partial_{t_1} ; \;\;\; 
\frac {\partial} {\partial q^i} = \epsilon \frac {\partial} {\partial q^i} ; \;\;\; 
\frac {\partial} {\partial v^i} = \epsilon \frac {\partial} {\partial v^i} \]
and
\[ N = N^{eq} + \epsilon N^{(1)} + \epsilon^2 N^{(2)} + \cdots \]
Here $\epsilon$ ($ << 1$) denotes a small number.  Thus, the Boltzmann equation (\ref{Boltzvv}) leads to the following two
equations,
\begin{equation}
\partial_{t_0} N^{eq} + \frac {\partial} {\partial q^i} (v^i N^{eq}) - \frac {\partial} {\partial v^i} (v^j v^k \Gamma^i_{jk} N^{eq}) 
= - \frac {1} {\tau} N^{(1)}
\label{aeq1}
\end{equation}
and 
\begin{equation}
\partial_{t_1} N^{eq} + \partial_{t_0} N^{(1)}
+ \frac {\partial} {\partial q^i} (v^i N^{(1)}) - \frac {\partial} {\partial v^i} (v^j v^k \Gamma^i_{jk} N^{(1)})  
= - \frac {1} {\tau} N^{(2)}
\label{aeq2}
\end{equation}
where, for simplicity, in the above we have used the BGK collision operator form $\Omega = - (N - N^{eq})/\tau$.\cite{BGK}
Taking the mass and momentum moments over (\ref{aeq1}), and use the properties in (\ref{EoS}) as well as conservation of mass
and momentum by the collision in (\ref{aconserv}), we immediately
obtain the leading (Euler) hydrodynamics as given by (\ref{amassconti}) and (\ref{aEuler2}),
\begin{eqnarray}
\partial_{t_0} \rho &+& \frac {1} {J} \frac {\partial} {\partial q^i} (J \rho u^i ) = 0
\nonumber \\
\partial_{t_0} (\rho u^i) &+& \frac {1} {J} \frac {\partial} {\partial q^j} (J \rho u^iu^j) 
+ \Gamma^i_{jk} \rho u^ju^k = - g^{ij} \frac {\partial p} {\partial q^j}
\label{aEuler2-1}
\end{eqnarray}
with an ideal gas equation of state, $p \equiv \rho\theta$.

Eqn.(\ref{aeq1}) can be inverted to give,
\begin{equation}
N^{(1)} = - \tau [\partial_{t_0} N^{eq} + \frac {\partial} {\partial q^i} (v^i N^{eq}) 
- \frac {\partial} {\partial v^i} (v^j v^k \Gamma^i_{jk} N^{eq})] 
\label{aeq1-invt}
\end{equation}
With the definition of particle density distribution function (\ref{adensity}), 
$N({q}, {\bar v}, t) = J({q}) f({q}, {\bar v}, t)$, and $J$ depends on ${q}$ only, 
hence eqn.(\ref{aeq1-invt}) is equivalent to
\begin{equation}
f^{(1)} = - \tau [\partial_{t_0} f^{eq} + \frac {1} {J} \frac {\partial} {\partial q^i} (v^i J f^{eq}) 
- \frac {\partial} {\partial v^i} (v^j v^k \Gamma^i_{jk} f^{eq})] 
\label{aeq1-invtf}
\end{equation}
With (\ref{aEuler2-1}), it is easily checked via straightforward algebra that $N^{(1)}$ (and $f^{(1)}$) gives
vanishing mass and momentum moments, namely
\[ \int d{\bar v} N^{(1)}({q}, {\bar v}, t) = 0, \;\;\; \int d{\bar v} v^i N^{(1)}({q}, {\bar v}, t) = 0 \]
On the other hand, taking the momentum flux moment, we have from (\ref{aeq1-invt}) the following,
\begin{eqnarray}
J \Pi^{ij,neq} &\equiv& \int d{\bar v} v^iv^j N^{(1)} 
\nonumber \\
&=& - \tau \int d{\bar v} v^iv^j [\partial_{t_0} N^{eq} 
+ \frac {\partial} {\partial q^k} (v^k N^{eq}) 
- \frac {\partial} {\partial v^k} (v^l v^m \Gamma^k_{lm} N^{eq})] 
\label{aneqflux0}
\end{eqnarray}
The last term in (\ref{aneqflux0}) above can be further simplified below,
\begin{eqnarray}
\tau \int d{\bar v} v^iv^j \frac {\partial} {\partial v^k} (v^l v^m \Gamma^k_{lm} N^{eq})
&=&  - \tau \int d{\bar v} v^lv^m \Gamma^k_{lm} N^{eq} \frac {\partial} {\partial v^k} (v^i v^j)
\nonumber \\
&=&  - \tau \int d{\bar v} v^lv^m (\Gamma^i_{lm} v^j + \Gamma^j_{lm} v^i) N^{eq}
\label{afluxlast}
\end{eqnarray}

Substituting (\ref{afluxlast}) into (\ref{aneqflux0}), we obtain
\begin{eqnarray}
J \Pi^{ij,neq} = &-& \tau \{ \int d{\bar v} v^iv^j [\partial_{t_0} N^{eq} 
+ \frac {\partial} {\partial q^k} (v^k N^{eq}) ] \\
\nonumber
&+& \int d{\bar v} v^lv^m (\Gamma^i_{lm} v^j + \Gamma^j_{lm} v^i) N^{eq} \}
\label{aneqflux}
\end{eqnarray}
Or equivalently, we have
\begin{equation}
J \Pi^{ij,neq} = - \tau \{ J\partial_{t_0} \Pi^{ij,eq} 
+ \frac {\partial} {\partial q^k} (J Q^{ijk,eq} )
+ J(\Gamma^i_{kl} Q^{jkl,eq} + \Gamma^j_{kl} Q^{ikl,eq}) \}
\label{flux-lbm}
\end{equation}
where the equilibrium heat flux $Q^{ijk,eq}$ is defined as
\begin{equation}
Q^{ijk, eq} = \int d{\bar v} v^iv^jv^k f^{eq} 
\label{aeqheat}
\end{equation}
The integration (\ref{aeqheat}) is straightforward to perform with $f^{eq}$ (given by (\ref{MaxBoltz})), and it yields
\begin{equation}
Q^{ijk, eq} = \rho\theta (g^{ij} u^k  + g^{jk} u^i  + g^{ki} u^j)  + \rho u^iu^ju^k
\label{aeqheatq}
\end{equation}
Therefore, together with (\ref{EoS}), eqn.(\ref{flux-lbm}) becomes
\begin{eqnarray}
J \Pi^{ij,neq} =
&-& \tau \{ J[g^{ij}\partial_{t_0} (\rho\theta ) + \partial_{t_0} (\rho u^iu^j)] 
\nonumber \\
& & + \frac {\partial} {\partial q^k} [J \rho\theta (g^{ij} u^k  + g^{jk} u^i  + g^{ki} u^j)  + J \rho u^iu^ju^k] 
\nonumber \\
& & + \Gamma^i_{kl} [J \rho\theta (g^{jk} u^l  + g^{kl} u^j  + g^{lj} u^k)  + J \rho u^ju^ku^l] 
\nonumber \\
& & + \Gamma^j_{kl} [J \rho\theta (g^{ik} u^l  + g^{kl} u^i  + g^{li} u^k)  + J \rho u^iu^ku^l]  \}
\label{aneqfluxq}
\end{eqnarray}

Next, let us take into account the properties of the metric tensor in the flat space, these are
\begin{eqnarray}
\frac {\partial} {\partial q^k} g^{ij} &+& \Gamma^i_{kl}g^{jl} + \Gamma^j_{kl}g^{il} = 0
\nonumber \\
\frac {\partial} {\partial q^k} (J g^{jk}) &+& J\Gamma^j_{kl}g^{kl} = 0
\nonumber \\
\frac {\partial} {\partial q^k} (J g^{ki}) &+& J\Gamma^i_{kl}g^{kl} = 0
\label{gamma-metric}
\end{eqnarray}
Then we can further simply (\ref{aneqfluxq}) to 
\begin{eqnarray}
J \Pi^{ij,neq} =
&-& \tau \{ J[g^{ij}\partial_{t_0} (\rho\theta ) + \partial_{t_0} (\rho u^iu^j)] 
\nonumber \\
&+& g^{ij}\frac {\partial} {\partial q^k}(J\rho \theta u^k) + J g^{jk}\frac {\partial} {\partial q^k}(\rho\theta u^i)
+ J g^{ki}\frac {\partial} {\partial q^k}(\rho\theta u^j) 
\nonumber \\
&+& \frac {\partial} {\partial q^k} (J \rho u^iu^ju^k) + \Gamma^i_{kl} J \rho\theta g^{jk} u^l + \Gamma^i_{kl} J \rho u^ju^ku^l 
\nonumber \\
&+& \Gamma^j_{kl} J \rho\theta g^{ki} u^l + \Gamma^j_{kl} J \rho u^iu^ku^l  \}
\label{aneqfluxq-g}
\end{eqnarray}

For the purpose of deriving hydrodynamics for isothermal fluid, we set $\theta = const$, in addition we substitute
the Euler order relations from (\ref{aEuler2-1}) for the terms with $\partial_{t_0}$, then after some straightforward
algebra, eqn.(\ref{aneqfluxq-g}) becomes
\begin{equation}
\Pi^{ij,neq} =
- \tau \rho\theta \{ g^{ik} (\frac {\partial u^j} {\partial q^k} + \Gamma^j_{kl}u^l) 
+ g^{jk} (\frac {\partial u^i} {\partial q^k} + \Gamma^i_{kl}u^l) \}
\label{aneqfluxq-final}
\end{equation}
Using the standard definition of covariant derivative in differential geometry, 
\[u^i|_k \equiv \frac {\partial u^i} {\partial q^k} + \Gamma^i_{kl}u^l \]
together with the rate of strain definition as the symmetric form of velocity derivative, 
\[ S^{ij} = S^{ji} \equiv \frac {1} {2} (g^{jk} u^i|_k + g^{ik} u^j|_k ) \]
then eqn.(\ref{aneqfluxq-final})
further simplies to
\begin{equation}
\Pi^{ij,neq} = - 2\mu S^{ij}
\label{aneqfluxq-final-s}
\end{equation}
where $\mu \equiv \tau \rho \theta$ is the dynamic viscosity.

Taking mass and momentum moments of eqn.(\ref{aeq2}), and the vanishing right-hand side, we obtain
\begin{eqnarray}
\partial_{t_1} \rho &=& 0
\nonumber \\
\partial_{t_1} (\rho u^i) &=& \frac {1} {J} \frac {\partial} {\partial q^j}(2J \mu S^{ij}) 
+ 2\Gamma^i_{jk}\mu S^{jk} 
\label{avisc}
\end{eqnarray}
Combining the first order hydrodynamics of (\ref{avisc}) with that of the Euler order (\ref{aEuler2-1}),
we finally arrive at the full Navier-Stokes hydrodynamics in a curvilinear coordinate system
\begin{eqnarray}
\partial_t \rho &+& \frac {1} {J} \frac {\partial} {\partial q^i} (J \rho u^i ) = 0
\nonumber \\
\partial_t (\rho u^i) &+& \frac {1} {J} \frac {\partial} {\partial q^j} (J \rho u^iu^j) 
+ \Gamma^i_{jk} \rho u^ju^k = - g^{ij} \frac {\partial p} {\partial q^j}
\nonumber \\
&+& \frac {1} {J} \frac {\partial} {\partial q^j}(2J \mu S^{ij}) 
+ 2\Gamma^i_{jk}\mu S^{jk} 
\label{aNS-1}
\end{eqnarray}
with an ideal gas equation of state, $p \equiv \rho\theta$.
Eqns.(\ref{aNS-1}) are in fact the mass continuity and the Navier-Stokes equation in coordinate-free operator forms
\begin{eqnarray}
\partial_t \rho &+& \nabla \cdot (\rho {\bf u} ) = 0
\nonumber \\
\partial_t (\rho {\bf u}) &+& \nabla \cdot (\rho {\bf u} {\bf u}) 
= - \nabla p + \nabla \cdot (2\mu {\bf S}) 
\label{acoord-free}
\end{eqnarray}

\section{Incompressibility in Curved Space with Underlying Euclidean Metrics}

The size of a volume element in a $3^2$ single particle phase-space is denoted as $J_p d{q} d{\bar v}$,
where $d{q} \equiv dq^1dq^2dq^3$ and $d{\bar v} \equiv dv^1dv^2dv^3$.  Clearly, for a curvilinear coordinate
system, $J_p = J^2 = g$, with $g \equiv det[g_{ij}]$ being the determinant of the metric tensor, and $J = J({q})$ is the Jacobian
of the curvilinear coordinates.

The transport of a particle phase-space density function $\phi = \phi ({q}, {\bar v}, t)$, without collision, 
follows a continuity equation in phase space, 
\begin{equation}
\partial_t (J_p \phi ) + \frac {\partial} {\partial q^i} (J_p \phi v^i) + \frac {\partial} {\partial v^i} (J_p \phi {\dot v}^i) = 0
\label{conti-vlasov}
\end{equation}
Since a particle moves along a straight path (geodesics) for the underlying Euclidean space according to the 1st law of Newton,
we have (see Appendix A)
\begin{equation}
{\dot v}^i = - \Gamma^i_{jk} v^j v^k
\label{straight-line}
\end{equation}
We can show that the motion of particles in phase-space is incompressible.  Define the velocity field
in phase-space as,
\begin{equation}
{\dot {\bf X}} \equiv ({\dot {q}}, {\dot {\bar v}}) = ({\bar v}, {\dot {\bar v}})
\label{phase-velocity}
\end{equation}
then, the divergence of the velocity field in phase-space is given by
\begin{equation}
\nabla \cdot {\dot {\bf X}} \equiv
\frac {\partial} {\partial q^i} (J_p v^i) + \frac {\partial} {\partial v^i} (J_p {\dot v}^i)
\label{div-v}
\end{equation}
Substitute (\ref{straight-line}) in eqn.(\ref{div-v}), we get
\begin{equation}
\nabla \cdot {\dot {\bf X}} =
\frac {\partial} {\partial q^i} (J_p v^i) - \frac {\partial} {\partial v^i} (J_p \Gamma^i_{jk} v^j v^k)
\label{div-v-g}
\end{equation}
But since both $J_p$ ($= J^2$) and $\Gamma^i_{jk}$ are functions of ${q}$ only, the second term in (\ref{div-v-g}) becomes
\begin{eqnarray}
\frac {\partial} {\partial v^i} (J_p \Gamma^i_{jk} v^j v^k) &=& J_p \Gamma^i_{jk} \frac {\partial} {\partial v^i} (v^j v^k) 
= J_p \Gamma^i_{jk} [v^j \frac {\partial v^k} {\partial v^i} + v^k \frac {\partial v^j} {\partial v^i}]
\nonumber \\
&=& J_p [\Gamma^i_{ji}v^j + \Gamma^i_{ik}v^k] = 2 J_p \Gamma^i_{ij}v^j
\label{div-v-k}
\end{eqnarray}
where the last equality in the above is due to the symmetry of the Christoffel symbol, $\Gamma^i_{jk} = \Gamma^i_{kj}$.
Furthermore, using the fundamental property 
\[ \Gamma^i_{ij} = \frac {1} {J} \frac {\partial J} {\partial q^j} \]
then (\ref{div-v-k}) becomes,
\begin{eqnarray}
\frac {\partial} {\partial v^i} (J_p \Gamma^i_{jk} v^j v^k) &=& 2 J_p \Gamma^i_{ij}v^j 
= 2 J \frac {\partial J} {\partial q^j}v^j 
\nonumber \\
&=& \frac {\partial J_p} {\partial q^j}v^j = \frac {\partial} {\partial q^j} (J_pv^j)
\label{div-v-h}
\end{eqnarray}
the last equality is because $\frac {\partial} {\partial q^j} v^j = 0$. 
Plug the result of (\ref{div-v-h}) into (\ref{div-v-g}), hence we have
proved the incompressibility property of the velocity field in the phase-space. That is,
\begin{equation}
\nabla \cdot {\dot {\bf X}} =
\frac {\partial} {\partial q^i} (J_p v^i) + \frac {\partial} {\partial v^i} (J_p {\dot v}^i) = 0
\label{incompressibility}
\end{equation}

With the incompressibility property of the phase space velocity field, 
the continuity equation (\ref{conti-vlasov}) takes on a form of the
Vlasov equation,
\begin{equation}
\partial_t \phi + v^i\frac {\partial \phi} {\partial q^i} + {\dot v}^i \frac {\partial \phi} {\partial v^i} = 0
\label{vlasov}
\end{equation}

It is worth to point out, the incompressibility property of the phase-space velocity field
is an intrinsic property of the particle motion via Newtonian mechanics in Euclidean
space.  Hence this is true in any coordinate systems whether Cartesian or curvilinear. 
Nonetheless, it is useful to directly show that such a property is preserved in a curvilinear coordinate system. 

Lastly, it is also convenient to define a density function below,
\[ f({q}, {\bar v}, t) \equiv J({q})\phi ({q}, {\bar v}, t) \]
so that hydrodynamic moments are given by simple integrations of ${\bar v}$, for instance
\[ \rho = \int d{\bar v} f = \int dv^1dv^2dv^3 f; \;\;\; \rho u^i = \int d{\bar v} v^i f = \int dv^1dv^2dv^3 v^i f \]

\end{document}